\documentclass[preprint, 3p, twocolumn]{elsarticle}
\usepackage{graphicx}
\usepackage{booktabs}
\usepackage{tabularx}
\usepackage[normalem]{ulem}
 \useunder{\uline}{\ul}{}
 \usepackage{rotating}
 \usepackage{dsfont}
 \usepackage{textcomp}
 \usepackage{upgreek}

\usepackage{color}
\def\Manu#1{\textcolor{black}{#1}}
\usepackage{tabularx}
\newcolumntype{Y}{>{\centering\arraybackslash}X}
\usepackage[switch]{lineno}
\usepackage{graphicx,xcolor}
\usepackage{epstopdf}
\usepackage{setspace}
\usepackage[T1]{fontenc}
\usepackage{amssymb,amsfonts,amsmath}
\usepackage{setspace}
\usepackage{subcaption}
\usepackage{mathtools}
\usepackage{etoolbox}
\biboptions{sort&compress}

\usepackage{titletoc}
\titlecontents{section}
[0pt]                                               
{}%
{\contentsmargin{0pt}                               
    \thecontentslabel\enspace%
    \normalsize}
{\contentsmargin{0pt}\large}                        
{\titlerule*[.5pc]{.}\contentspage}                 

\patchcmd{\maketitle}
  {\finalMaketitle}
  {\finalMaketitle\tableofcontents\vspace{\baselineskip}}
  {}{}

\usepackage{hyperref}
\hypersetup{
    colorlinks=true,
    linkcolor=blue,
    filecolor=magenta,      
    urlcolor=cyan,
}
\begin{document}

\title{\Manu{Foaming and antifoaming in non-aqueous liquids}}
\author[vinny,equal]{S.G.K. Calhoun}
\author[vinny,harvard,equal]{V. Chandran Suja}
\ead{vinny@stanford.com}

\author[vinny]{G.G. Fuller}\ead{ggf@stanford.com}
\address[vinny]{Department of Chemical Engineering, Stanford University, Stanford, California 94305, USA}
\address[harvard]{School of Engineering and Applied Sciences, Harvard University, Massachusetts 01234, USA}
\address[equal]{Equal contribution}

\begin{abstract}




Investigation into the physics of foaming has traditionally been focused on aqueous systems. Non-aqueous foams, by contrast, are not well understood, but have been the subject of a recent surge in interest motivated by the need to manage foaming across industrial applications. In this review, we provide a comprehensive discussion of the current state-of-the-art methods for characterizing non-aqueous foams, with a critical evaluation of the advantages and limitations of each. Subsequently we present a concise overview of the current understanding of the mechanisms and methods used for stabilizing and destabilizing non-aqueous foams. We conclude the review by discussing open questions to guide future investigations.


\end{abstract}

\begin{keyword}
   Non-aqueous Foams,  Lubricants, Diesel fuels, Cooking Oils, Interferometry, Air entrainment/release, Defoaming/Antifoams, 
\end{keyword}

\maketitle

\section{Introduction}\label{sec:Introduction}
    
    
Foams are a dispersion of a gas in a solid or a liquid phase. Liquid foams are ubiquitous, and are further classified as either aqueous or non-aqueous. Aqueous foams, where the liquid is water, have been extensively studied with wide industrial applications from detergents to food science, and cosmetics to fire-fighting \cite{garrett2016science, pugh1996foaming}. When the continuous liquid is not water, the foam is refered to as a non-aqueous foam. Unlike the plethora of studies on aqueous foams, non-aqueous foams are not well understood, and have only begun to be intensively studied in the last decade \cite{friberg2010foams,blazquez2014non}. They are, however, commonly encountered in a number of applications: lubricating oil systems such as in engines or wind turbines; molten polymers; the fossil fuel industry; manufacturing; and the food industry \cite{garrett2016science,pugh2016bubble,frostad2016coalescence,suja2018evaporation,Schmidt1996nonchap,suja2020foam}. Understanding the mechanisms involved and tuning the stability of non-aqueous foams is crucial for the above-mentioned applications. In some instances stable foam is desired as part of a product, such as in a number of food products. In other instances foams are detrimental and need to destabilized such as those encountered in liquid filling operations and lubrication. In both these cases, non-aqueous foams present unique challenges not found in aqueous foams.
    
Foams are thermodynamically unstable due their large interfacial area (and hence high interfacial free energy). In the absence of kinetic or thermodynamic stabilization mechanisms to maintain their improbable voluminous structure, foams are evanescent such as in the case of pure liquids like water. Common kinetic stabilization mechanisms include viscous stabilization, the Marangoni effect, and interfacial rheology driven stabilization \cite{koczo2017lubricants}.  Kinetic stabilization in aqueous foams is typically due to the last two mechanisms originating from the adsorption of a surface active species to the air-water interface \cite{binks2010non}. Aqueous systems have high air-liquid surface tensions ($\sim 72\;mN/m$), encouraging surfactants to adsorb strongly to the interface. Non-aqueous systems have surface tensions much lower than that of aqueous systems ($\sim 15-30 \; mN/m$), which makes adsorption thermodynamically unfavorable for most surface active species \cite{binks2010non,fameau2017nonEdible,blazquez2014non}. This is a major aspect that differentiates non-aqueous systems from aqueous systems. In addition, thermodynamic stabilization mechanisms such as disjoining pressure are also less prevalent in non-aqueous systems due to weak repulsion in electrostatic double layers and low dielectric constants \cite{fameau2017nonEdible,blazquez2014non}. In lieu of these well known primary stabilizing modes, foaming is sustained by unique physio-chemical characteristics of non-aqueous systems including high viscosity, high thermal conductivity, species heterogeneity and differential volatility \cite{fameau2017nonEdible,dyab2013particle, suja2018evaporation}. These important differences make non-aqueous foams sometimes challenging to produce, and equally difficult to control through conventional antifoaming methods. 
    

In the rest of the manuscript we provide a succinct and comprehensive review of foaming and antifoaming mechanisms in non-aqueous systems. In section \ref{sec:Methods}, we outline and contrast the three common experimental platforms used to study non-aqueous foams - bulk foam tests, single film tests and single bubble tests.  In sections \ref{sec:Stabilization} and \ref{sec:Destabilization}, we provide a detailed overview of the current scientific understanding of non-aqueous foam stabilization and destabilization mechanisms respectively. Finally in section \ref{sec:Conclusion} we conclude the review and discuss important directions for future research.

\section{Methods for characterizing non-aqueous foams}\label{sec:Methods}

\subsection{Bulk Foam Tests}
The traditional and most popular method for evaluating non-aqueous foams is bulk foam tests (Fig.\ref{fig:ExperimentalSetup}A). Bulk foam tests recreate foams at a bulk scale, often in a manner that is similar to their natural origin. Quantification of foam stability is usually obtained by measuring the foam volume over time both during and after foam generation. The former quantity is a measure of the capacity to foam under the tested condition, while the latter is a direct measure of foam stability. Some of the common bulk foam tests are detailed below: 

\subsubsection*{Foam Rise Test}
One of the earliest and most common bulk foam test, the foam rise test consists of injecting air into a column of test liquid and measuring the generated foam volume over time (Fig.\ref{fig:ExperimentalSetup}A). The foam rise test mimics foam generation due to sparging and air released from surface reactions. An attractive feature of the foam rise test is the ability to precisely control the amount of air injected into the liquid. This allows the measurement of additional foam stability metrics such as the air-release fraction of the liquid. This test is commonly used to study non-aqueous lubricant \cite{binks2010non, suja2018evaporation, suja2020foam, tamai1978relation}, crude oil \cite{poindexter2002factors,chen2019foaming, callaghan1989non, bauget2001dynamic}, hydrocarbon fuel and edible oil foams \cite{mellema2004importance}.
\subsubsection*{Shake Test}
One of the simplest bulk foam tests, the shake test consists of shaking a container with a known volume of liquid for fixed duration. The generated foam volume is measured at end of the test to establish foam stability. Automated shake test rigs are commonly used to improve test reliability \cite{garrett1994experimental,denkov2004mechanisms} and eliminate variability arising from the shaking method \cite{rudnick2017lubricant}.  Shake foam tests best mimic foam generation during sloshing associated with transport of liquid, and is commonly used for qualitative evaluation of foaming in a number of situations \cite{hamdi1993surfactant,garrett1994experimental}.
\subsubsection*{Ross-Miles Test}
The Ross-Miles test, originally conceived by John Ross and Gilbert Miles in 1941 \cite{ross1941apparatus}, consists of generating foam by pouring/spraying a test liquid from a fixed height into a cylindrical container containing the same liquid. Foam stability is established by monitoring the height of foam over time. Originally developed to study foamability of detergents, Ross-Miles test best mimics foam generation associated with filling operations, and is  used to evaluate non-aqueous foams encountered during the handling of lubricants \cite{rudnick2017lubricant}, hydrocarbon fuels \cite{quigley2004biodiesel, grabowski1996efficient} and others \cite{egan1984properties}.
\subsubsection*{Mechanical Agitation Test}
As the name suggests, mechanical agitation tests generate foams by agitating liquids with machinery such as gears, mixer blades, perforated discs, and running engines \cite{rudnick2017lubricant}. If designed appropriately, these tests can faithfully mimic foam generation in the required application setting by accounting for effects of shear, air entrainment, antifoam breakdown and temperature - yielding practically accurate foamability and foam stability data. Due to the inability to independently control these physical effects, mechanical agitation tests are not commonly used for fundamental studies. Common examples include the Flender foam test where foams are generated by counter-rotating spur gears immersed in a test lubricant \cite{flenderfoamISO}, and engine oil aeration tests \cite{astm2013D6894standard} conducted inside internal combustion engines. 
 
\subsubsection*{Depressurization Foam Test} 
Depressurization foam tests evaluate foams generated by dissolved gases in liquid using a pressurized cell \cite{sheng1997experimental, blazquez2016crude}. During an experiment, step and linear decompression profiles are applied to a test liquid saturated with gases, and the evolution of the foam height is monitored to evaluate foam stability. Depressurization foam tests best mimic foam generated in conditions where the dynamic pressure in a liquid changes significantly. Such conditions are commonly encountered during lubrication and crude oil processing \cite{sheng1997experimental,zhang2021effects,blazquez2017crude}.

\begin{figure*}[!h]
\includegraphics[width=\linewidth]{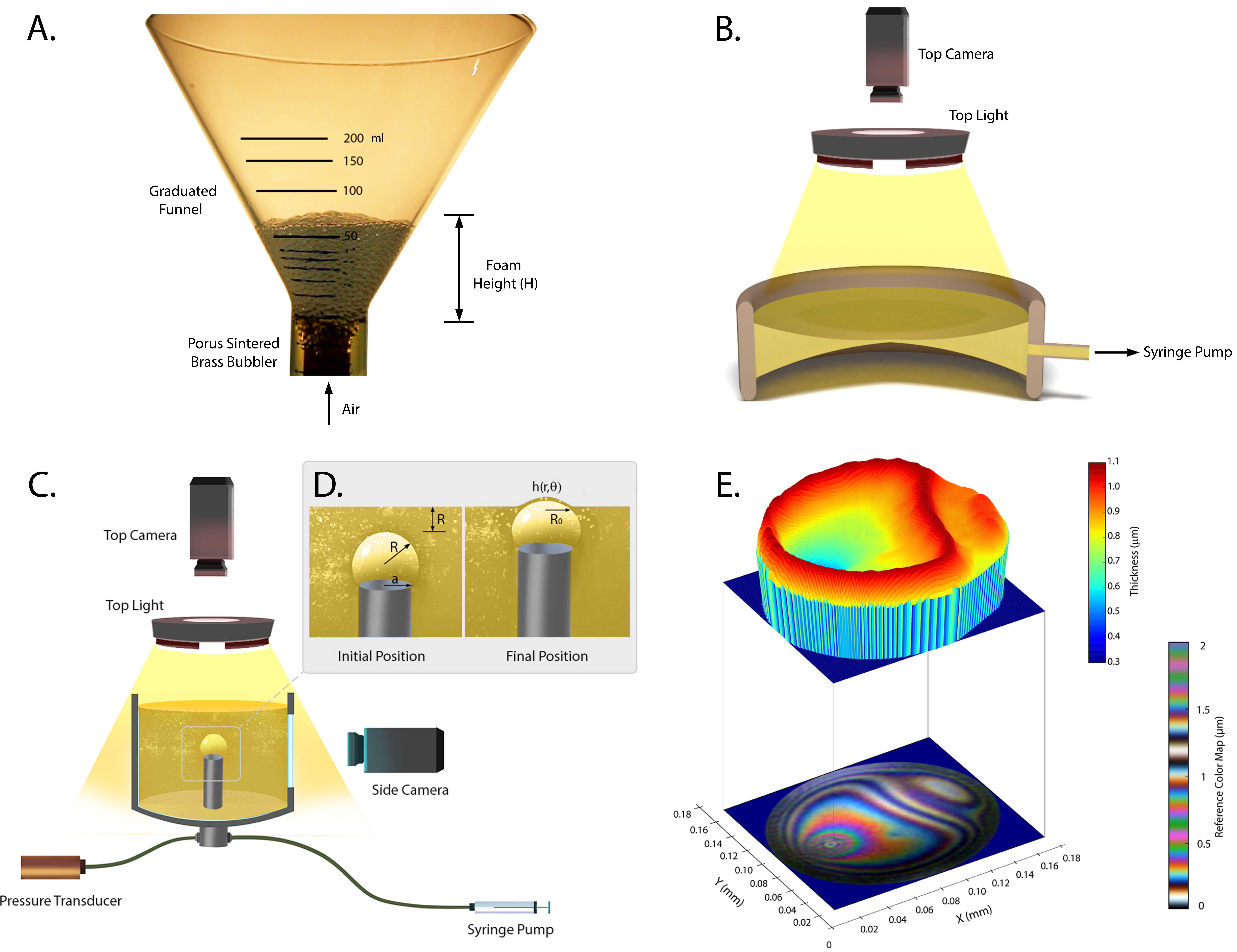}
\caption{Common experimental setups used for evaluating non-aqueous foams. {\bf a.} Bulk foam setup - Foam rise test {\bf b.} Single film setup -  Scheludko Cell {\bf c.} Single bubble setup - Dynamic Fluid Interferometer (DFI) {\bf. d} Bubble positions at different stages of a DFI experiment {\bf e.} Interferogram obtained using the DFI in a lubricant with an antifoam (bottom), and its film thickness reconstruction (top)}
\label{fig:ExperimentalSetup}
\end{figure*} 

\subsubsection*{Standardized Foam Tests}
A number of the above foam tests has been modified and adapted as standard foam tests by organizations such as the American Society for Testing and Materials (ASTM) and the International Organization for Standardization (ISO). ASTM D892 \cite{astm2013D892standard, centers1993behavior}, ASTM D6082 \cite{astm2017D6082standard} and ISO 6247 test \cite{iso6247standard} for evaluating lubricant foaming,  Japan Industrial Standard (JIS)-K2518 \cite{jisK2518standard} for evaluating petroleum foaming, and ASTM D7840 \cite{astm2017D7840standard} test for evaluating engine coolant foaming  are modified foam rise tests. ASTM D1173 \cite{astm2015D1173standard} test, Deutsche Industrie Norm (DIN)-53902 \cite{din53902standard}, and French Norm (NF)-M07-075 (also referred to as the BNPE test) \cite{nfM07075} for evaluating diesel fuel foams, are modified Ross Miles tests. ASTM D6894 \cite{astm2013D6894standard} and D8047 \cite{astm2019D8047standard} for evaluating engine oil foaming, and ISO 12152 \cite{iso6247standard} for evaluating gear oil foaming are mechanical agitation tests.
 
 \subsubsection*{Advantages and Limitations}
 Bulk foam tests best mimic real life foams, and capture all their complexity including many body interactions, the effects of advection, Ostwald ripening and the presence of plateau borders. They are easily adopted to simulate the natural origin of foam, are relatively easy to perform, and allow for the convenient measurement of aggregate foam properties. As a result, bulk foam tests have become the standard way of testing foam stability in industrial settings.
 
 Despite their advantages, these tests are not suitable to systematically probe stabilization mechanisms due to the complexity of bulk foams. Simplifications like 2D foams do exist \cite{miralles2014foam}, however, these systems are still inconvenient for developing a detailed understanding of the stability of thin liquid films that ultimately dictate foam stability. To overcome the limitations of bulk tests, researchers have developed other complementary methods. 

\subsection{Single Film Tests}
Tests based on single films - the simplest abstraction of foams - evaluate the stability of free standing liquid films analogous to those formed when two bubbles approach each other within a foam \cite{exerowa2018foam}. Perhaps the most well known single film setup is the Scheludko-Exerowa cell, which was originally developed by Derjaguin and subsequently improved by Scheludko, Exerowa and Mysels \cite{scheludko1959device, mysels1964soap, sheludko1967thin,exerowa1997foam}. Other common variants include the Exerowa-Scheludko porous plate cell, the Mysels cell, the bike-wheel microcell  and the Dippanear cell \cite{exerowa2008emulsion, mysels1966direct, pereira2001bike, dippenaar1982destabilization, garrett2016science}.  Detailed reviews on single film setups and results are available in the literature \cite{exerowa1997foam, langevin2015bubble}.

A typical single film setup, the Scheludko-Exerowa cell, is shown in Fig.\ref{fig:ExperimentalSetup}B. The Scheludko-Exerowa cell consists of a cylindrical cell for holding a test liquid, a bore in the side of a cell connected to a pump for controlling liquid volume (not shown in figure), pressure transducer, and arrangements for generating and recording thin film interferograms. The test cell is usually placed inside a sealed container (not shown in figure) to minimize external disturbances and minimize evaporation. During a typical experiment, the cylindrical cell is filled with a test liquid. The liquid is gradually withdrawn from the cell through a bore in side of the cylinder, creating a circular thin film that expands as the liquid is withdrawn. The withdrawal is stopped after the film reaches a desired radius. Subsequently, the film drains under a constant capillary pressure generated at the plateau border \cite{exerowa2018foam, zhang2015domain}. The dynamics associated with film thinning are monitored by the simultaneous recording of thin film interferograms and the pressure in the film. Both the non-aqueous film stabilization mechanisms and film coalescence time are interpreted by processing the obtained interferograms and pressure traces \cite{sheludko1967thin,andersson2010disjoining,bergeron1997thin}. 

 \subsubsection*{Advantages and Limitations}
Single film experiments have transformed our understanding of thin film stability. In particular, due to the ability of the technique to measure the pressure within the film, we have a deep understanding of the role of disjoining pressure in terminal thin film drainage and thin film stability \cite{stubenrauch2003disjoining}. Further, single film results have also aided in improving the theoretical understanding of thin film drainage, as the symmetric films produced in these experiments \cite{exerowa2018foam} have made them attractive for theoretical and numerical analyses \cite{joye1994asymmetric}.

Despite the above advantages, single film experiments have certain limitations including difficulties in conveniently controlling the size of the film and the approach velocity of the interacting interfaces, and the inability to study the interaction of interfaces with different radii of curvature.

\subsection{Single Bubble Tests}
Single bubble tests evaluate the stability of a bubble interacting with either another bubble or a flat liquid-air interface \cite{suja2020single}. In terms of mimicking real life foams, single bubble tests fall midway between bulk foam tests and single film tests. The earliest reported single bubble experiments can be traced back to works of William Bate Hardy \cite{hardy1912tension} and James Dewar \cite{dewar1919soap}. More practical versions of single bubble/drop setups can be seen in the works of Rehbinder and Wenstrom \cite{rehbinder1930stabilisierende}, while a comprehensive version of a single bubble setup, consisting of an arrangement to form bubbles in a controlled way along with an interferometry setup for measuring the film thickness, can be found in a 1960 publication by Stanley Mason  \cite{charles1960coalescence}.

A typical schematic of a single bubble is shown in Fig.\ref{fig:ExperimentalSetup}C. A single bubble setup commonly consists of a chamber to contain the test liquid, a capillary, and a syringe pump to form the bubble \cite{suja2020single}. In many cases, a pressure transducer is also connected to the capillary for monitoring the bubble pressure \cite{chandran2016impact}. The bubble profile is visualized by a side camera, while the spatiotemporal evolution of the liquid between the bubble and the air-liquid interface is visualized by the top camera. At the start of a typical experiment, the chamber is filled with a test liquid and a bubble formed. The experiment starts by moving the bubble at a fixed velocity towards the air-liquid interface from its initial to its predetermined final position (Fig.\ref{fig:ExperimentalSetup}D). The final position of the bubble is usually selected such that it corresponds to the equilibrium position attained by a free bubble through the balance of buoyancy and capillary forces. Concurrently, the top camera records the spatiotemporal evolution of the liquid between the bubble and the air-liquid interface, while the pressure transducer monitors the pressure inside the bubble. Similar to the single film experiments, film stabilization mechanisms and film coalescence time are interpreted by processing the obtained interferograms (see Fig.\ref{fig:ExperimentalSetup}E) and pressure traces \cite{suja2018evaporation,suja2020foam,suja2020hyperspectral,suja2020symmetry}. Other variants of the single bubble experiments include those that probe freely rising single bubble interactions with an air-liquid interface \cite{mckendrick1991thin, menesses2019surfactant}, and those that probe bubble-bubble interaction between separate bubbles immobilized on capillaries \cite{paulsen2014coalescence}.

\subsubsection*{Advantages and Limitations}
 Single bubble/drop experiments have three notable advantages over single film tests.  Firstly, single bubble/drop experiments allow the use of a complete bubble/drop, thus enabling the effects of the dispersed phase size and the rise velocity \cite{frostad2016coalescence} to be independently studied.  Secondly, single bubble/drop experiments can probe the interaction of interfaces with different radii of curvature, and have notably improved the understanding of coalescence at flat liquid-air interfaces \cite{suja2018evaporation,suja2020foam,kannan2018monoclonal}. Thirdly, in situ interfacial dilatational rheology measurements can be conveniently performed in single bubble/drop setups, thus making them a more holistic tool for developing a  mechanistic understanding of thin film stability \cite{kannan2018monoclonal}.

The short comings of single bubble tests include the inability to measure the pressure within the liquid film for directly probing disjoining forces in the film, and their unsuitability for probing foaming in translucent and opaque liquids such as crude oils.

\section{Non-aqueous Foam Stabilization Methods and Mechanisms} \label{sec:Stabilization}
Non-aqueous foams exist, and exhibit a range of persistence across applications, due to a variety of mechanisms that stabilize the foams. Some mechanisms are similar to those observed in aqueous foams, but others are more specifically present in or particularly important to non-aqueous foams.

\begin{figure*}[!h]
\includegraphics[width=\linewidth]{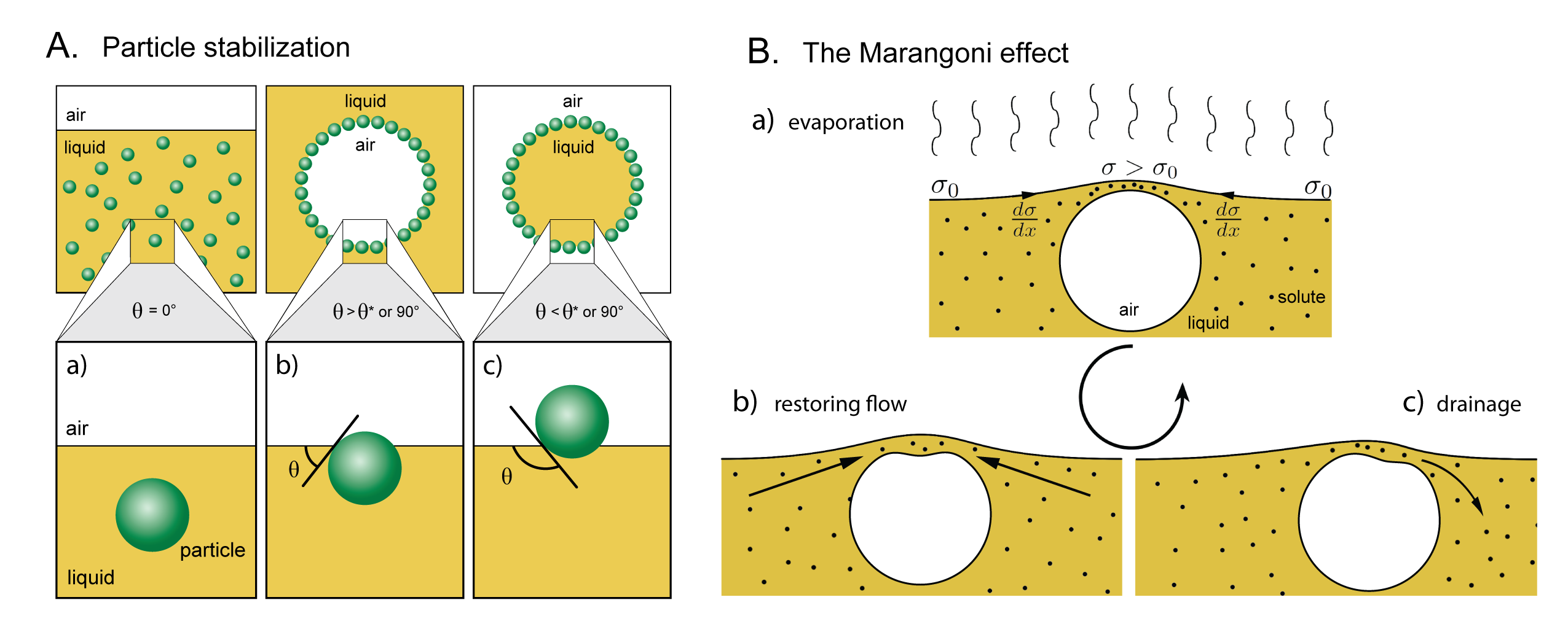}
\caption{Common stabilization mechanisms of action in non-aqueous foams. {\bf A.} Particle stabilization: a) complete particle wetting with $\theta=0$\textdegree, b) Stable foams with partial particle wetting $\theta< 90$\textdegree, c) Stable emulsions with minimal particle wetting $90$\textdegree$< \theta< 180$\textdegree. {\bf B.} Evaporation-induced stabilization, illustrating the Marangoni effect: a) Evaporation causes a surface tension gradient between bulk and thin film, b) Restoring flow into the thin film from the bulk, c) Drainage \& repeat}
\label{fig:Stabilization}
\end{figure*}

\subsection{Particle stabilization} \label{subsec:Particle}
Colloidal particles can adsorb to the air-liquid interface and stabilize foams through particle-interface and particle-particle interactions \cite{hunter2008role}. By interacting with the interface and removing an area of open interface, adsorption of a particle at the interface can reduce total free energy of the system \cite{fameau2017non,binks2011particles,binks2002solid,binks2021review}.  Particle-particle interactions also retard foam collapse through steric interactions. This is well known and frequently occurs in aqueous foams, and can also occur in non-aqueous systems. Adsorption will occur if the particles have an intermediate wettability or hygrophilicity. Hygrophilicity, a non-aqueous corollary to hydrophilicity, means an affinity for liquids \cite{murakami2010particle}. One way to measure this quantity is through the solid-liquid-air contact angle $\theta$ (Fig. \ref{fig:Stabilization}A). In non-aqueous systems, hygrophilic particles are those that are completely wettable by the non-aqueous liquid ($\theta \sim 0$\textdegree) and thus will remain dispersed in the bulk and not adsorb to the interface (Fig. \ref{fig:Stabilization}A.a). Intermediately hygrophilic particles (Fig. \ref{fig:Stabilization}A.b, $0 $\textdegree$ < \theta < 90$\textdegree) adsorb to the interface with the majority of the particle within the liquid and are effective in stabilizing foams. Meanwhile, relatively hygrophobic particles (Fig. \ref{fig:Stabilization}A.c, $90$\textdegree$< \theta< 180$\textdegree) adsorb to the interface with the majority of the particle within the gaseous phase, and are therefore effective in stabilizing liquid-in-air droplets \cite{binks2021review, binks2011particles,  fameau2017nonEdible, garrett2018anti}. Details of particle stabilization including adsorption energy, stratification, electrostatic repulsion, capillary forces and self-assembly have been investigated especially by Kralchevsky et al \cite{danov2010multipoles,kralchevsky2016repulsion,kralchevsky1990interaction}.

Particle stabilization is far more difficult in non-aqueous foams because oils completely wet most particles, even with controlled surface chemistry \cite{murakami2010particle}. This is due to the relatively low surface tension of oils as compared to aqueous systems, which makes it far more difficult to find or produce hygrophobic particles that can spontaneously adsorb to the interface. To the degree it is possible, tuning the wettability and contact angle for a particular system is controlled through either surfactant adsorption onto the particle surface or chemical modification of the particle surface \cite{fameau2017nonEdible}. Unlike in aqueous foams where any hydrophobic particles suffice, this is challenging in non-aqueous foams due to the need to find chemical groups with a lower surface tension than the oils in the system. In addition, particles should be partially hygrophobic with contact angles between 0 and 180 degrees, eliminating the possibility of particles surfaced by both hydrocarbon-containing groups, which are fully wetted by most oils, and fluoro groups, which can be fully hygrophobic to some oils \cite{fameau2017non,murakami2010particle,binks2006}. With a contact angle of approximately $50$\textdegree, n-hexadecane on a smooth surface of poly(tetrafluoroethylene) does form stable foams, though is far from the hygrophobic thershold to form liquid-in-air droplets \cite{murakami2010particle}. Surface roughness encourages hygrophobicity, thus giving another degree of freedom in particle design. Another method to optimize contact angle and wettability of the particles is by changing the liquid phase surface tension by choosing a different base oil or using an oil mixture \cite{binks2010non,binks2011particles,binks2010particles,murakami2010particle}. 
    
Due to the wettability and low surface tension challenges with non-aqueous systems, stabilization via particles is not particularly common. Examples of particle stabilized non-aqueous foams include dichlorodimethylsilane (DCDMS)-modified amorphous silica or organo-modified Laponite clay nanoparticle stabilized glycerine or ethylene glycol foams\cite{dyab2013particle}, and OTFE particle (agglomerated oligomers of tetrafluoroethylene particle) stabilized perfluorostyrene/divinylbenzene mixture foams \cite{murakami2010particle}.  Silica nanoparticles are also known to strongly stabilize crude oil foams above a critical hydrophobicity \cite{chen2019foaming}. Further information on particle stabilization can be found in reviews by Fameau and Saint-Jalmes \cite{fameau2017non}, and Binks et al \cite{binks2021review}. Applications of particle based foam stabilization in food science and other non-petroleum industry applications can be found in a recent review by Fameau et al \cite{fameau2020food}. 


\subsection{Surfactant stabilization} \label{subsec:Surfactant} 
Surfactants (surface active small molecules) can impart both thermodynamic and kinetic stabilization to foams by adsorbing to the interface. Thermodynamically, adsorption of surfactant molecules to the air-liquid interface reduces the surface tension and lowers the total free energy of the system. Surfactants can also impart disjoining pressures within films and retard foam collapse. Kinetically, surfactants can stabilize foams through Maragnoni stresses (see Sec.\ref{subsec:Marangoni}) and steric effects. Quite commonly used in aqueous foams, surfactants can sometimes stabilize non-aqueous foams, contingent on the molecular structure of the surfactant. This is much more rare than in aqueous systems, because the magnitude of reduction in surface tension that surfactants can produce is low due to the decreased absolute surface tension of non-aqueous systems \cite{koczo2017lubricants}.  

Stabilization or destabilization of a foam can be tuned by choice of surfactant type and concentration. Hydrocarbon-based surfactants generally do not adsorb to the interface and thus lead to unstable non-aqueous foams, but in some instances can produce very stable foams in conditions where the surfactant becomes insoluble and form crystalline particles at the interface \cite{fameau2017nonEdible,fameau2017non,blazquez2014non,pugh2016bubble}. Crystalline particles and the resulting oleogels are especially applicable to edible foams including those made from fatty acids, fatty alcohols, mixtures of mono and diglycerides, triacylglycerols or sucrose esters and sunflower lecithin \cite{fameau2017nonEdible,shrestha2006crys,fameau2015crys,brun2015crys,binks2016}. Polymethylsiloxane-based surfactants can produce stable non-aqueous foams due to the lower surface tension from incompatibility of the various molecular groups with each other and the surrounding non-aqueous liquid; along with this formation of a surface tension gradient, these surfactants also increase surface viscosity, promoting foams \cite{fameau2017nonEdible,friberg2010foams,blazquez2014non}. The surfactant category that most significantly lowers surface tension (thereby stabilizing non-aqueous foams) is the fluoroalkyl surfactants, due to the fluorine atom’s covalent radius being the proper size to shield the carbon chain and decrease interactions \cite{fameau2017nonEdible,blazquez2014non}. A final group of common surfactants that can stabilize non-aqueous foams are asphaltenes and resins, which are found in or added to crude oil and affect the amount of foaming through both their concentration and molecular weight \cite{fameau2017nonEdible,blazquez2014non,blazquez2016crude,friberg2010foams, rodriguez2020asphaltene}. Further details on the principles of surfactant stabilization can be found in a recent review article by Manikantan and Squires \cite{manikantan2020surfactant}, and more details on edible foams are available in two recent reviews by Fameau et al \cite{fameau2017non} and Heymans et al \cite{heymans2017crys}. 
    
\subsection{Solutocapillary Marangoni stabilization}\label{subsec:Marangoni}
The Marangoni effect can stabilize foams by retarding the thinning and rupture of foam lamellae. This effect arises when an induced surface tension gradient drives the flow of the sub-phase liquid from an area of lower surface tension to an area of higher surface tension. When the gradient of surface tension results from spatial inhomogeneities in  concentration, such as that of a solute (see Sec.\ref{subsec:Evaporation}) or of a surfactant, the effect is termed as solutocapillary Marangoni \cite{rodriguez2019evaporation,pugh1996foaming}. In foams, heterogeneity in surfactant concentration can occur due to drainage. The viscous stresses from foam drainage drags surfactants in the direction of flow, creating a gradient in surfactant concentration. The resulting gradient in surface tension induces counteracting Marangoni flows that usually pull the sub-phase liquid back toward the center of the film \cite{nierstrasz1999marginal, suja2018evaporation, callaghan1989non,marangoni1972principle, mellema2004importance,poulain2018ageing}. This restoring flow slows drainage several fold compared to a surfactant-free surface \cite{koczo2017lubricants}, thus increasing the lifetime of the thin film and thereby stabilizing the foam. Strength of the solutocapillary Marangoni effect can be optimized by controlling surfactant concentration and chemistry. Solutocapillary Marangoni foam stabilization is much more prevalent in aqueous systems than non-aqueous foams.  Further details on Marangoni stabilization and thin film dynamics is available in a recent review by Chatzigiannakis et al \cite{chatzigiannakis2021thinfilm}. 
    
\subsection{Evaporation-induced stabilization}\label{subsec:Evaporation}
Evaporation can drive a special case of solutocapillary Marangoni flows, thereby counter-intuitively stabilizing non-aqueous foams (Fig. \ref{fig:Stabilization}B). Many non aqueous systems, particularly petroleum products are mixtures of liquids, each having different surface tensions and volatilities. Evaporation of the volatile liquid components can generate spatial heterogeneities in species concentration due to spatial variations in foam lamellae thickness. The largest changes in species concentration occurs in the thinnest portion of the film, which is usually at the center of the foam lamellae. If the residual non-volatile liquid components in the film have a higher surface tension,  the evaporation-induced concentration gradients drive solutocapillary Marangoni flows that pull liquid from the bulk into the thinning film, stabilizing the film and thus the entire foam \cite{suja2018evaporation,suja2020symmetry,rodriguez2019evaporation, callaghan1989non}. 

This type of stabilization depends on the differential volatilities and surface tension of component liquids, and is therefore especially important for non-aqueous foams with their heterogeneous composition and wide range of volatilities. Foam stabilization due to this mechanism can be easily controlled by changing the relative volume fraction of the various liquids and by varying component volatilities\cite{rodriguez2019evaporation,suja2018evaporation}. Evaporation-induced foam stabilization is common in lubricating oils and diesel fuels that have a large number of component liquids with large variations in volatilities among these components \cite{suja2018evaporation}. Further details on this stabilization method is available in studies reported by Rodriguez et al \cite{rodriguez2019evaporation} and Chandran Suja et al \cite{suja2018evaporation}.
    
\subsection{Thermal Marangoni effect}
In addition to concentration gradients, the Marangoni effect can also occur due to temperature gradients. In most liquids, surface tension changes inversely with temperature. Thus temperature gradients can create surface tension gradients that drive flows from areas of higher temperature to areas of lower temperature. These temperature gradients can be due to heating or a consequence of other processes \cite{callaghan1989non}.  Evaporation is one such process that can cause decreases in temperature, with larger drops in temperature occurring in the thinnest part of the film \cite{menesses2019surfactant}. Evaporation-induced thermal Marangoni stabilization is particularly salient in liquids with high volatility and low specific heats \cite{suja2020single}. Non-uniform heating, and Rayleigh–B{\'e}nard convection currents arising in liquid heated from below (or cooled from above) are other common sources of thermal gradients. Heating induced thermal Marangoni stabilization is common, for example, in cooking oils \cite{mellema2004importance}.

\subsection{Phase separation based stabilization} 
In addition to thermal and evaporation driven Marangoni flows, phase separation phenomena can also set up restoring Marangoni flows that stabilize foams in certain situations. Mixtures of partially miscible liquids show maximum foaming and foam stability when the composition approaches a liquid-liquid phase separation boundary \cite{ross1977relation,pugh2016bubble,ross1975bitri,ross1996proanti,ross1981dynamic}. Decreasing solubility of dissolved components in the solvent causes solute to move to the interface as phase separation occurs - adsorbing similarly to surfactants. The phase separation and subsequent adsorption increase surface active material at the interface and can generate surface tension gradients. These surface tension gradients induce Marangoni flows, stabilizing the individual thin film and thus the foam. Tuning of this stabilization mechanism can be accomplished by the addition of components that will affect the solubility, and thus the phase separation boundary. In addition, changing the temperature of the system can lead to changes in the phase separation behavior, and thus the stabilization and foaming propensity. 


\subsection{Viscous stabilization}\label{subsec:Viscosity}
Viscosity is an important influence to consider in non-aqueous foam stabilization, as viscosity varies by many orders of magnitude across common non-aqueous liquids. Film drainage rate decreases with increasing viscosity, stabilizing and increasing the lifetime of the foam, and vice versa \cite{callaghan1989non,binks2010non}. Indeed, without any added foam-stabilizing components, liquids with low bulk viscosity generally cannot form foams at all as the films rupture very quickly. This is a common and straightforward stabilization mechanism, and can be easily tuned to control foam stability, particularly in non-aqueous foams. Viscosity modifying chemicals, namely those that increase viscosity, can be added to a system to stabilize non-aqueous foams. In addition to stabilizing foams, high viscosity also makes generating foams difficult, which can lead to a non-monotonic relationship between viscosity and foaming. Many lubricating oils have both high viscosity and a variety of stabilizing additives such as particles, or surface active surfactants or polymers, leading to rather stable foams \cite{koczo2017lubricants}. In a related manner, in certain systems such as some crude oil foams, temperature variation can indirectly affect foam stability by influencing viscosity, namely via an increasing temperature decreasing the viscosity and thereby decreasing foaming \cite{callaghan1989non}.

\subsection{Interfacial rheology}
Similar to aqueous systems \cite{kannan2018monoclonal,tammaro2021flowering, kamkar2020polymeric}, interfacial rheological properties including surface shear and dilatational viscoelasticity can stabilize non-aqueous thin films as well \cite{callaghan1989non}. Surface shear viscosity, similar to bulk viscosity, can stabilize foams by retarding drainage \cite{Schmidt1996nonchap,langevin2015bubble,langevin2000influence}. Surface dilatational viscosity and dilatational elasticity are also known to increase foam stability by resisting deformation of the lamellae and retarding film thinning \cite{callaghan1983inter, callaghan1989non}. Interfacial rheology stabilized foams are common in food science, such as the so-called oleofoams that are stabilized by interfacial elasticity arising from the interfacial adsorption of crystalline aggregates of surfactants, fatty acids, or waxes (also see Section \ref{subsec:Surfactant}) \cite{fameau2017nonEdible}. Interfacial rheology stabilization can be tuned through choosing surface active materials with different molecular weights, structures and changing concentrations. This method of stabilization is ripe for further exploration especially outside food science applications. 

\section{Non-aqueous Foam Destabilization Methods and Mechanisms} \label{sec:Destabilization} 
Non-aqueous foams are in some cases desired and promoted (with the methods above), while in other cases, particularly in lubricating oils and petroleum industries, they are detrimental. Presence of foam in place of an oil can make engine operation inefficient, obstruct fueling or bottling, cause wear on gears, retard crude oil processing, or precipitate excess heat and damage to wind turbines. Fortunately, non-aqueous foams can be destabilized by a variety of methods. One common way to control foams is through the use of chemical additives called antifoams that deploy a variety of foam destabilization mechanisms to destroy foams. Once an antifoam is added, determining the mechanism of action at play for a particular system is indispensable for improving antifoam efficiency.

Along with the paucity of work on non-aqueous foams in general, mechanisms of destabilization specific to non-aqueous foams have seen especially little direct investigation. A number of existing foam stabilization mechanisms are therefore conjectured based on analogies from findings for aqueous foams, and on knowledge of the properties of non-aqueous systems. Here we will discuss solid, liquid, and mixed types of antifoam additives and the current consensus of the various destabilization mechanisms of action they utilize. We conclude this section by discussing two non-conventional antifoaming methods, namely acoustic and mechanical antifoaming.


\subsection{Solid antifoams}
\label{subsec:Solid}
One physical manifestation of antifoam additives is in the form of solid particles added to the bulk liquid \cite{garrett2015defoaming}. It is well known that solid antifoams typically utilize the bridging-dewetting mechanism of destabilization (Fig. \ref{fig:Destabilization}A) \cite{denkov2014mech}. In aqueous foams, these particles are required to be hydrophobic, while in non-aqueous foams this is not required \cite{garrett2018anti}. Ease of entry into the liquid-air interface is also an important aspect of solid antifoam effectiveness, see Section \ref{subsec:Liquid} for discussion of entry barrier and thermodynamic entry coefficient.

\subsubsection{Bridging-dewetting mechanism}
\begin{figure*}[!h]
\includegraphics[width=\linewidth]{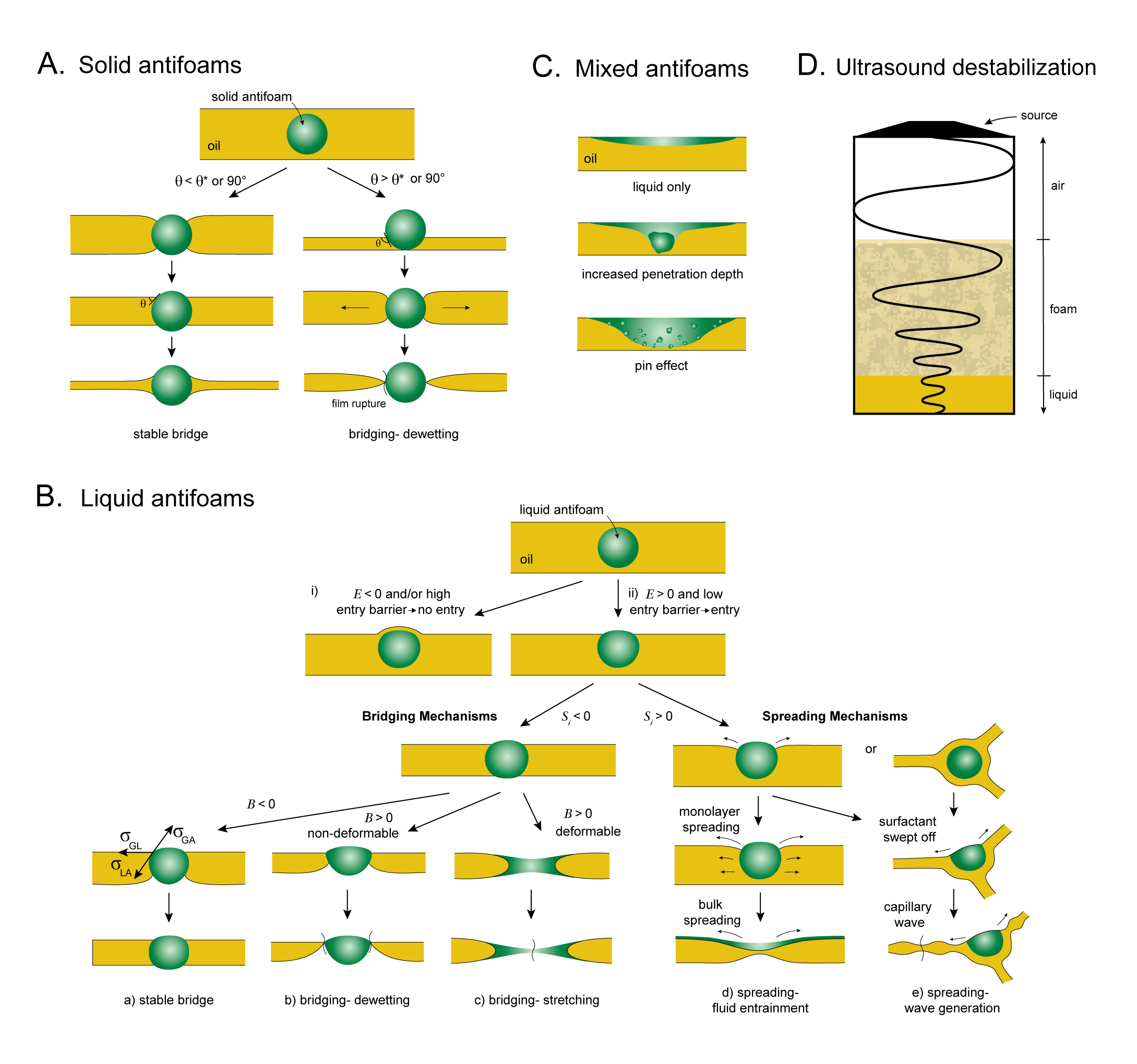}
\caption{Common destabilization mechanisms of action in non-aqueous foams. {\bf A.} Solid antifoams. {\bf B.} Liquid antifoams, including entry $E$, bridging $B$, and spreading $S$ coefficients: i) Unsuccessful entry, ii) Entry. Mechanisms: a) Stable Bridge, b) Bridging- Dewetting, c) Bridging- Stretching, d) Spreading- Fluid entrainment, and e) Spreading- Wave generation mechanisms. {\bf C.} Mixed antifoams, showing increased penetration depth and pin effect with added solids. {\bf D.} Ultrasound destabilization. }
\label{fig:Destabilization}
\end{figure*}

In the bridging-dewetting mechanism, an antifoam particle located within a thinning foam film comes into contact with the two opposite liquid-air interfaces, forming a bridge between them. The liquid then dewets off the antifoam surface, perforating at the solid-liquid-air contact point and rupturing the film (Fig. \ref{fig:Destabilization}A).  Degree of hygrophobicity of the solid particles is the primary determining factor of antifoam efficiency by the bridging-dewetting mechanism \cite{murakami2010particle,denkov2004mechanisms}. This particle hygrophobicity can be measured with the solid-liquid-air contact angle $\theta$. If the particles are sufficiently hygrophobic, with the contact angle exceeding a critical value $\theta^*$, this solid bridge dewets from the liquid leading to rupture of the thin film. On the other hand, if $\theta < \theta^*$, the foam film remains stable. This situation can even enhance foam stability, as discussed in section \ref{subsec:Particle} \cite{denkov2004mechanisms,frye1989part,garrett2016science,denkov2014mech}.  This critical value of contact angle for dewetting of the particle is often found to be 90 \textdegree for smooth convex surface particles, commonly used particles such as spheres and ellipses. Solid particles with sharp edges and corners can dewet at lower contact angles \cite{garrett2015defoaming,dippenaar1982destabilization}. Further details of particle destabilization have been discussed by Kralchevsky et al \cite{kralchevsky2001foams}.

\subsection{Liquid antifoams} \label{subsec:Liquid}
Antifoams can also be purely liquid, which disperse in the base liquid as small droplets. Liquid antifoams are the predominant type of antifoam used in applications with non-aqueous foams. Effective antifoam liquids are characterized by lower surface tensions than those of the foaming liquid, and are insoluble \cite{garrett2018anti}. Viscosity and concentration of liquid antifoams also influence their effectiveness  \cite{garrett2015defoaming}. 


Liquid antifoam droplets destabilize foams by: becoming lenses on the air-liquid interface, spreading on the interface, and/or bridging the two interfaces of the film (Fig. \ref{fig:Destabilization}B). The precise mechanism of destabilization depend on a set of coefficients that are dictated by the interfacial tensions of the system (of the base liquid, antifoam globule, and air). An antifoam globule must first enter into the liquid-air interface for an antifoam to have an opportunity to function. The thermodynamic aspect of drop entry is determined by the entry coefficient $E$ :
\[ E = \sigma_{GL} +\sigma_{AL} -\sigma_{AG} \]
where $\sigma_{GL}$, $\sigma_{AL}$, $\sigma_{AG}$ are the interfacial tensions of the gas (air)-foaming liquid, antifoam-foaming liquid, and antifoam-gas (air) interfaces, respectively \cite{kralchevsky2001foams}. Positive values of $E$ show an affinity of the drop for the interface, while globules with negative values of $E$ remain wetted and immersed in the bulk liquid (Fig. \ref{fig:Destabilization}B.i-ii). A bridging coefficient $B$:
\[ B = \sigma_{Gl}^2 +\sigma_{AL}^2 -\sigma_{AG}^2 \] 
correlates with the stability of a bridged antifoam and thus likelihood of rupture, with negative $B$ connected with stable bridges and vice versa \cite{denkov1999bridging, garrett2018anti}. Positive values of $B$ necessarily signify $E$ is positive as well, though the reverse is not true \cite{denkov1999bridging, garrett2015defoaming}. The spreading coefficient $S$:
\[ S = \sigma_{GL} -\sigma_{AL} -\sigma_{AG} \]
measures the tendency of the antifoam globule to spread at the air-liquid interface, with positive values indicating an affinity for spreading. For each coefficient, it is frequently fruitful to differentiate between an initial value (e.g. $S_i$) and an equilibrium value (e.g. $S_{eq}$), utilizing $\sigma_{GL}^i$ in absence of spread oil antifoam on the surface and $\sigma_{GL}^{eq}$ in the presence of spread oil, respectively\cite{garrett2018anti}. In addition, it is usually more appropriate, in calculating these coefficients $E$ $S$ and $B$, to use dynamic surface tensions in place of equilibrium surface tensions, as we examine destabilization in systems with dynamic foams. 

The entry barrier is an equally important consideration along with the $E$, $S$ and $B$ coefficients in determining whether an antifoam will be effective. The entry barrier encompasses the kinetic aspect of drop entry and accounts for contributions from solution viscosity and long range electrostatic forces that hinder drop entry. A low entry barrier is required for any antifoaming ability, while a high entry barrier will forestall even oils with highly positive coefficients \cite{denkov2014mech}. Adding hygrophobized solid particles into the oil to form a mixed antifoam is the most efficient way to reduce the entry barrier \cite{denkov2014mech, denkov1999bridging, garrett2018anti}. Further details on $E$, $S$, $B$, and the entry barrier, and their implications for the destabilization mechanisms and antifoam effectiveness, can be found in Denkov et al \cite{denkov2014mech} and Garrett \cite{garrett2018anti}.

\subsubsection{Bridging Mechanisms}
Formation of an antifoam bridge between the two interfaces of a foam film is a possibility with liquid antifoams post droplet entry (Fig. \ref{fig:Destabilization}B.a-c). The bridging coefficient $B$ discussed previously encapsulates the stability of a bridged antifoam and thus its tendency to rupture a film. Positive $B$ values correspond with unstable bridges and film rupture, while negative $B$ values are associated with stable bridges that do not rupture the film (Fig. \ref{fig:Destabilization}B.a) \cite{kralchevsky2001foams,garrett2018anti}. Unstable bridges can proceed to rupture a film through potential ensuing processes such as the bridging-stretching or the bridging-dewetting mechanism.  

\subsubsection*{Bridging-Stretching mechanism}
 In the bridging-stretching mechanism, after the bridge forms it develops into a biconcave shape with the thinnest region in the bridge center \cite{garrett2015defoaming, garrett2018anti,denkov1999bridging}. The bridge then stretches radially due to unbalanced capillary pressures at the antifoam-foaming liquid and antifoam-air interfaces. In time this stretching results in a thin unstable antifoam oil film that ruptures at the center, severing the foam lamella as well (Fig.\ref{fig:Destabilization}B.c). This is generally a very fast mechanism of rupture, acting in the early stages of film thinning \cite{denkov2004mechanisms}. Observational evidence has confirmed occurrence of this mechanism, for example in certain systems with silicone oil antifoams  \cite{denkov2014mech,denkov1999bridging}. Fig.\ref{fig:ExperimentalSetup}E also shows direct experimental measurements of a system hypothesized to rupture via the bridging-stretching mechanism. The hydrodynamics of this process, however, has not been addressed through simulation, and many open questions remain, e.g. the roles of oil viscosity, $\theta*$, and the fate of the droplet after film rupture \cite{garrett2018anti}. Detailed knowledge of these variables will be useful for optimizing this mechanism for specific systems. 

\subsubsection*{Bridging-Dewetting Mechanism}
The bridging-dewetting mechanism discussed in the solid antifoam section (Section \ref{subsec:Solid}) is proposed to also occur with liquid antifoams, provided the oil droplet is sufficiently hygrophobic relative to the surrounding phase \cite{denkov1999bridging,denkov2014mech}. Here that would manifest as a liquid antifoam droplet bridging across the thin film and dewetting as the liquid pulls off the antifoam droplet, rupturing the film at the air-antifoam-foaming liquid contact point (Fig. \ref{fig:Destabilization}B.b). As deformability of antifoam globules makes it easier to achieve appropriate contact angles, in principle liquid and deformable mixed solid-liquid antifoams (see Section \ref{subsec:Mixed}) are more likely to dewet than with solid particles. However, evidence documenting the occurrence of the bridging-dewetting mechanism in liquid antifoams is not forthcoming in literature; documentation of this mechanism is perhaps difficult due to the experimental challenge of capturing the high speed phenomenon \cite{denkov2004mechanisms,dippenaar1982destabilization}. 

Use of this mechanism in non-aqueous systems could require more deliberate design to guarantee dewetting between two oils. For systems with positive bridging coefficients and thus unstable oil bridges, the relative rates of dewetting and deforming differentiate between the bridging-stretching mechanism and the bridging-dewetting mechanism. If the time scale of deformability relative to that of dewetting is slow, the film ruptures via dewetting \cite{denkov2004mechanisms}.

\subsubsection{Spreading Mechanisms}
Positive values of the spreading coefficient $S$ indicate a tendency for the antifoam globule to spread on the air-liquid interface. The presence of an oil film on the surface does not destabilize the thin film. The rupture promotion of spreading behavior could be due to induced sub-phase flows or due to removal of the surfactants stabilizing the interface \cite{pugh1996foaming,kralchevsky2001foams}. The two potential destabilization mechanisms that depend on spreading are discussed below and illustrated in (Fig. \ref{fig:Destabilization}B.d-e). Further discussion of spreading can be found in Chapter 3 of Garrett \cite{garrett2016science}.  

\subsubsection*{Spreading-Fluid Entrainment Mechanism}
 The spreading-fluid entrainment mechanism is proposed to occur when an antifoam globule spreads rapidly on the foam film surface and pulls the sub-phase liquid along with it, leading to accelerated film thinning and rupture (Fig.\ref{fig:Destabilization}B.d). Modification of the velocity/degree of antifoam spreading and ratio of surface tensions between antifoam and foaming liquid allow for optimization of performance. The existence of this phenomenon as a cause of rupture is disputed, with a lack of supporting evidence especially in non-aqueous foams. Nonetheless, there is evidence of spreading-fluid entrainment having an enhancing effect on other mechanisms of destabilization \cite{denkov2004mechanisms,denkov1999bridging}. 

\subsubsection*{Spreading-Wave Generation Mechanism} 
In the spreading-wave generation mechanism (Fig. \ref{fig:Destabilization}B.e), an antifoam globule in a plateau border spreads on the surface and in so doing, partially sweeps aside adsorbed surfactants that dampen surface capillary waves through Gibbs elasticity. Resulting areas with reduced density of adsorbed surfactant are both less stabilized by the surfactant and are prone to the appearance of capillary waves. The combination of these effects create a susceptibility to local thinning and rupture, sometimes despite a large thickness averaged across the film \cite{denkov2014mech, denkov2004mechanisms, garrett1993anti,aveyard1993break}. Optimization can be undertaken by controlling the surfactant concentrations, Gibbs elasticity, and relative surface tensions of the foaming liquid and antifoam. This mechanism has been documented in aqueous foams; however, perhaps due to the rarity of surfactant-stabilized systems in non-aqueous foams, there is inadequate evidence of its effectiveness in these systems, meriting further studies. 

\subsection{Mixed antifoams} \label{subsec:Mixed}
Mixtures of solid hydrophobic particles with oils, called antifoam compounds or mixed antifoams, is another construct of antifoam that is commonly used. The addition of the solid component to an oil facilitates the entry of the antifoam globule into the air-liquid interface by reducing the entry barrier (discussed in Section \ref{subsec:Liquid}) \cite{denkov2014mech}. This entry barrier reduction occurs due to the "pin effect", in which sharp edges of the particle more easily make direct contact with the film surface due to lower repulsive electrostatic forces (Fig.\ref{fig:Destabilization}C) \cite{denkov2004mechanisms,garrett1994experimental,garrett2015defoaming,garrett2018anti}. Solid particles included in mixed antifoams also improve performance by increasing the “penetration depth” over a solely liquid antifoam (Fig.\ref{fig:Destabilization}C). Oil lenses from liquid antifoams extend through a certain fraction of the film thickness and form a bridge when the lens contacts the opposite surface of the film; added solid particles increase this fraction of the film thickness, accelerating bridging \cite{denkov2004mechanisms,denkov1999bridging,garrett2016science}.  

Mixed antifoams utilize many of the same mechanisms as the liquid antifoams. Most mechanisms of action are based upon the instability of configurations with antifoams bridging across thin films or plateau borders \cite{garrett2018anti}. As with liquid antifoams, mixed antifoams can rupture films via the bridging-stretching mechanism and via the bridging-dewetting mechanism. Mixed antifoams can be rigid or deformable in shape, depending on exact composition of the antifoam. Bridging-stretching requires that the globule is deformable, while mixed antifoam globules with limited deformability are more likely to employ a bridging-dewetting mechanism \cite{denkov2004mechanisms}. 

Current understanding is that the solid and liquid components of a mixed antifoam have a synergistic effect because they play complementary roles; the role of the solid particles is to destabilize the oil-antifoam-air films through the pin effect and greater penetration depth, while the oil component lends deformability to the globules and spreads on the interface. This balance between the roles of oil and particles is essential for the proper functioning of the antifoam. A disturbance of this balance can lead to a deactivation of mixed antifoams, with a deterioration in performance over time, when the mixture gradually segregates into particle-rich and particle-free globules \cite{denkov2014mech}. This balance is also an opportunity to tune the properties of an antifoam as desired for a specific system. In aqueous systems, mixed antifoams have been found to have the expected synergistic effect, and is more effective than either component alone. Thus mixed antifoams are the predominant antifoam type in aqueous antifoams. However, in non-aqueous systems, mixed antifoams are currently less common \cite{garrett2018anti}. Some non-aqueous applications do currently draw upon mixed antifoams and could likely see more development in the future. 


\subsection{Viscosity modifiers} 
Viscosity is an important foam stabilizer (see Sec.\ref{subsec:Viscosity}), and consequently viscosity modifiers are effective in destabilizing foams.  A first category of viscosity modifiers is any chemical that decreases the bulk viscosity and accelerates film drainage and eventual rupture.  Another type of effective viscosity modifier is a chemical that decreases the surface viscosity and surface elasticity, promoting rupture by removing some of the stabilizing mechanisms discussed in section \ref{sec:Stabilization}. This change usually occurs through competitive adsorption of a new molecule to the interface that gives the interface lower surface viscosity and elasticity. Rupture promotion with a new liquid added in this way can also be due to reduction in concentration of surfactants stabilizing the interface, reducing the surface tension and removing the surface tension gradient \cite{pugh1996foaming}.

\subsection{Ultrasound destabilization}
Ultrasound as a method of both antifoaming (preventing foaming) and defoaming (destructing an existing foam) has been around for nearly seventy years. In this method, an ultrasonic generator, often piezoelectric, is placed in the air above the foam, and the waves traveling outward defoam existing foams or act as an antifoam preventing foam generation (Fig. \ref{fig:Destabilization}D). If the source is placed in the liquid itself, the relatively high velocity of sound in these liquids leads there to be little transfer of ultrasonic energy from the liquid into the foam, and thus this configuration is ineffective on a foam.

The mechanisms responsible for ultrasound based foam destabilization is still not well studied or understood, particularly in the non-aqueous foam context \cite{garrett2015defoaming}. Foam destruction by ultrasound is usually attributed to the induced acoustic pressure, the resonant vibration in bubbles, turbulence produced by acoustic pressure and enhanced drainage from induced capillary waves \cite{gallego2015ultra}.   The efficacy of ultrasound antifoaming/defoaming can be optimized by adjusting the acoustic intensity and power \cite{pugh1996foaming}. In addition to tuning acoustic intensity and power, frequency is also a variable that can be optimized, as there is some evidence that resonant vibrations of bubbles in foam can encourage defoaming \cite{garrett2015defoaming,pugh1996foaming}. In some cases, such as old foams, ultrasound vibrations can also have a stabilizing and foam-boosting effect, but it is unclear when stabilizing or destabilizing will occur; combining these unknown specifics with the general lack of understanding of ultrasound in non-aqueous foams, many open questions remain. As ultrasound based foam destabilization is non-invasive and non-contaminating,  this method is commonly employed to control aqueous and non-aqueous foams encountered during food processing \cite{mawson2016airborne}.



\subsection{Mechanical destabilization}
Chemical modes of controlling foams, such as the solid, liquid and mixed antifoams discussed above, are large in number and widely effective. However, in certain situations such as in some food products, chemical defoaming methods are infeasible, especially due to concerns of product contamination. Mechanical methods of foam destabilization are unfettered by problems such as these, and are thus worthy of consideration.  Commonly utilized mechanical methods include rupturing bubbles through mechanical shock generated by shear, compressive, impact or centrifugal forces, pressure changes, or suction. Frequently used devices include vacuum chambers, liquid or air jet streams, cyclones, and rotary devices. It is worth noting that these methods are sometimes more costly than chemical antifoams and defoamers, and can still contaminate a product unless operable in sterile conditions. They also become challenging in large scale industrial applications, as devices are complex and expensive, and scaling up raises problems \cite{gallego2015ultra,deshpande2000mechanical}. Still, mechanical techniques of destabilization present viable alternatives to chemical methods, and can be well utilized in specific applications if not more widely.


\section{Conclusions}\label{sec:Conclusion}
Non-aqueous foams are widely useful in a variety of industries and it is fortunate that recent years have hosted an explosion of interest and investigation that has considerably improved our understanding of this neglected phenomenon. The primary goal of this review was to present the current state of research in non-aqueous foams, spanning an exploration of methods of characterization and testing,  mechanisms found to stabilize non-aqueous foams, and potential destabilization mechanisms that can be employed for antifoaming or defoaming. 

Due to fundamental differences in the physical properties of the liquids involved, non-aqueous and aqueous systems display differences in foam stabilization and destabilization mechanisms. Non-aqueous systems generally exhibit much lower surface tensions ($\sim 15-30 \; mN/m$) than aqueous ($\sim 72\;mN/m$), reducing surface adsorption, and much less electrostatic repulsion and low dielectric constants. As a result, unlike aqueous systems, surfactant-based foam stabilization is not common in non-aqueous systems. Instead, non-aqueous foams are commonly stabilized by mechanisms that leverage their unique physical properties such as high viscosity (viscous stabilization), heterogeneous composition (phase separation based stabilization, particle based stabilization),  differential volatility (evaporation induced solutocapillary stabilization), and high thermal conductivity (thermocapillary stabilization). Foam destabilization mechanisms in non-aqueous systems are less well studied and are often inferred from aqueous literature. Well established non-aqueous foam destabilization mechanisms include bridging-dewetting in solid antifoam particles and bridging-stretching in liquid antifoams.

Despite the recent progress in understanding of non-aqueous foams, many open questions remain unaddressed. Firstly, considering foam stabilization, there are outstanding questions in the mechanisms of crystalline particle based foam stabilization. The role of size, shape, polymeric state and solution rheological properties of crystalline particles on foam stability is not well understood. Evaporation induced foam stabilization is also relatively less explored outside lubricants. The role of this mechanism in driving foams in systems such as fuels and in multicomponent volatile liquids close to phase separation is unknown. Secondly, considering foam destabilization,  a full understanding of the processes resulting from antifoam bridge instability, and the route toward rupture, is still elusive. The bridging-stretching mechanism is one of those processes, but is likely not  the sole process that exists in response to an unstable oil bridge. Further simulation and experimental work is required, to prove existence of the proposed bridging-dewetting process and identify other potential processes \cite{garrett2018anti}. Spreading based foam destabilization mechanisms are also poorly understood. Existence of the proposed spreading-fluid entrainment mechanism of antifoam is disputed in general, while evidence of the spreading-wave generation mechanism in non-aqueous foams specifically is also insufficient. 
Answering these questions, and thus improving our understanding and optimization, will start to unlock the full potential 
of non-aqueous foams.

\bibliographystyle{vancouver-annote}
\bibliography{Reference}
\end{document}